# Critical Amplitude Fluctuations in Low Superfluid Density Two Dimensional Superconductors

J. A. Chervenak and J. M. Valles, Jr.
*Department of Physics, Brown University 02912*

**ABSTRACT**

We report results of transport measurements in the quantum critical regime of the disorder tuned, 2D superconductor-insulator transition (SIT) in homogeneously disordered films. We show that, as the superconducting transition temperature decreases, the transition width grows, appearing to diverge at the SIT. In addition, structure develops in the DC current-voltage characteristics of films closest to the SIT indicating that the 2D superconductivity is driven into a regime of extreme inhomogeneity. The data suggest a picture of the phase transition in which large amplitude fluctuations occur as the amplitude is suppressed to near zero by disorder.

Increasing the normal state sheet resistance, $R_N$, of a superconducting film degrades its superconducting properties [1]. The size of the thermodynamic fluctuation dominated portion of the phase transition grows, both above [2] and below [3] the mean field transition temperature, $T_{c0}$, and $T_{c0}$ and quasiparticle lifetimes decrease [1,4]. As long as $R_N << h/4e^2 = R_Q = 6.45$ k$\Omega$, these effects can be understood with theories that assume that the low temperature state is a mean field BCS superconductor. As $R_N$ approaches $h/4e^2$, this assumption breaks down and quantum fluctuation effects are expected to appear. This regime has received a great deal of attention as it corresponds to where a transition from a superconducting to insulating phase occurs [5-9]. Numerous experiments suggest that quantum critical fluctuations associated with the superconductor-insulator quantum critical point dominate the finite temperature properties of films in this regime.

The fluctuations near a quantum critical point sample the phases in its immediate vicinity, and therefore relate directly to the physical origin of the phase transition. Previous studies of the 2D Superconductor-Insulator Transition (SIT) have focused on a proposed phase diagram in which increasing disorder in 2D superconductors induces an insulating phase consisting of localized Cooper pairs [10]. In this picture, fluctuations in the relative order parameter phases of the islands of Cooper pairs dominate fluctuations of the order parameter amplitude in the quantum critical regime. Experimentally, these *phase* fluctuations are manifest in granular films [9] and Josephson junction arrays [6] through the persistence of dissipation to temperatures far below the superconducting transition temperature, $T_{c0}$, at which Cooper pairs or the *amplitude* of the order parameter forms on the grains.

The applicability of this model to the behavior of homogeneously disordered films has been controversial. Previous electron transport and tunneling experiments on these systems indicate that a reduction in $T_{c0}$ [1,7,11], the superconducting energy gap $\Delta_0$ [12], and the condensation energy [8] are the predominant qualitative effects in the destruction of their superconducting state. This evolution contrasts strongly with the granular film case, for which $\Delta_0$ remains constant through the SIT [13]. Consequently, it has been suggested that the order parameter amplitude becomes so small near the SIT that it is likely to fluctuate [8,12]. Recent theories support this notion [14-16]. For example, a theory of the superconductor to normal metal quantum phase transition predicts that spatial fluctuations in $T_{c0}$ and $\Delta$ grow upon approaching this QPT [14]. The model correlates disorder induced broadening of the superconducting transition with the suppression of $\Delta_0$.

In this paper, we examine transport in ultrathin homogeneously disordered films tuned through the SIT. We show that the width of the portion of the resistance transitions dominated by amplitude fluctuations diverges at the SIT. In addition, the current-voltage, I(V), characteristics of films with the broadest transitions exhibit structure indicating the presence of multiple critical currents within a single film. These observations provide direct evidence of the growth of amplitude fluctuations near the SIT in homogeneous films. Moreover, they suggest that amplitude fluctuation effects dominate in the experimentally accessible region near the SIT in homogeneous films.

Samples were quench condensed onto a fire polished glass slide with predeposited Au/Ge contacts. At the substrate temperature $T_{glass}$ < 8K .8 nm of Sb was deposited.

Continuity of the Bi toplayer was typically detected at ~.55 nm, about two monolayers coverage. The films of interest occur where the crossover from insulator to superconductor is observed, between the film thicknesses of .7-.9 nm. Four terminal measurements on two or more different squares of film showed that the samples exhibited homogeneity of $R_{sq}$ and $T_{c0}$ throughout the film. $R_{sq}$ is the sheet resistance and $T_{c0}$ is the superconducting transition temperature taken where $R(T=T_{c0}) = R_N(T_{c0})/2$. $R_N(T)$, the temperature dependent normal state of the superconducting film, is obtained below. Depositions and measurements are performed on a sample stage in contact with a dilution refrigerator ($.05 < T < 10$ K) and centered in a superconducting magnet ($0 < H < 8$ T). AC resistance measurements were taken with a lock-in amplifier at low frequency (<10 Hz) in the regime where the voltage in the sample was linear with excitation current. DC IV measurements were digitized with filtered electronics kept outside the shielded room and verified by measurements taken with analog electronics and battery operated power sources and amplifiers.

Figure 1(a) shows the resistive transitions, $R(T)$, of a series of homogeneously disordered Bi/Sb films in which the electronic system is driven from its insulating phase to its superconducting phase by incremental thickness changes. As shown previously, $T_{c0}$ decreases continuously toward zero with increasing disorder on the superconducting side of the transition [7]. Simultaneously, the transitions become broader.

This broadening reflects the growth of the critical regime of the superconducting transition. And, since it occurs in the part of the transitions where $R$ first begins to drop, we interpret it as a growth of the paraconductance fluctuation dominated portion of the transition [2]. The original Aslamasov-Larkin theory predicts that the paraconductance fluctuation effects, a consequence of thermal fluctuations of the order parameter amplitude, grow with increasing $R_N$. The theory is in qualitative agreement with these data and in quantitative agreement with data on films far from the SIT [1]. Also, tunneling experiments on films with $T_{c0}$ as low as 1.8 K and $R_N \approx 4 k\Omega$ demonstrate that the energy gap opens near $T_{c0}$ [17,18]. This result indicates that the order parameter amplitude is small and likely to fluctuate through the part of the transition near $T_{c0}$. This interpretation is central to the analysis that follows and, as $T_{c0}$ is suppressed to near zero, it is fundamentally different from the interpretation that has led to "scaling" analyses of data and comparison to boson localization models of the SIT [5,7,10].

Within this interpretation, the observed transition widths are slightly accentuated by $R(T)$ increasing with decreasing temperature. To eliminate this normal state effect and measure the growth of the critical regime, it is necessary to normalize $R(T)$ with the normal state resistance, $R_N(T)$. We employed two different methods to determine $R_N(T)$ that yielded equivalent results (see later discussion). We show only one of them here. For films in the weakly localized regime, the increase of $R_N(T)$ with decreasing temperature results from negative corrections to the film conductance, $\delta G_{QC}(T)$, brought about by weak localization and disorder enhanced electron-electron interaction effects. These corrections show little variation with sheet resistance for $R_N < 40\ k\Omega$ [19,20]. Replotting the data of Fig. 1(a) as the change in the conductance of each of the films relative to its conductance at 8K, $\Delta G(T)$, in Fig. 1(b) illustrates this point. The $\Delta G(T)$ converge to the same temperature dependence at high temperatures where superconducting fluctuations are not evident [8]. At low temperatures, the $\Delta G(T)$ of all the films, save one, increase strongly as they become superconducting. The one exception is closest to or on the insulating side of the SIT and we define it to be the normal state conductance, $\delta G_{QC}(T)$. The superconducting fluctuation contribution to the transport of each of the other films is given by $\Delta G_{sc} = \Delta G(T) - \delta G_{QC}(T)$. And, the normalized resistances can be calculated using:

$$\frac{R(T)}{R_N(T)} = 1 - \Delta G_{sc} R(T) \qquad (1)$$

The results of applying this procedure to the data of Figs. 1a and b are shown in Fig.1c. The data in Fig. 1c indicate that superconducting fluctuations occur over an increasingly large temperature range as the SIT is approached. In the most extreme case, for the film closest to the SIT, for which $T_{c0}<.06$ K, $R/R_N$ has dropped to $<.95$ at 1K indicating the presence of fluctuations at temperatures more than a factor of 20 above $T_{c0}$.

To quantify how the size of the amplitude fluctuation dominated region varies with disorder, we measure the temperature interval over which $R/R_N$ drops from 0.9 to 0.1. We show this transition width, $\Delta T/T_{c0}$, for Bi/Sb over the range $.2R_Q<R_N \sim R_Q$ and $.05<T_{c0}/T_{cbulk}<.4$ plotted against the logarithm of $T_{cbulk}/T_{c0}$ in Fig. 2. In each case, $\Delta T/T_{c0}$ grows substantially as the SIT is approached and, for our definition of $\Delta T$, exceeds 1 for films closest to the SIT. It appears to diverge logarithmically [21]. The same trend was indicated in a series of PbBi/Ge films over a narrower $T_{c0}$ range.

Structure appears at low temperatures in the current-voltage (IV) characteristics of the films with the broadest transitions ($T_{c0}<0.4$ K). Figure 3 shows the IV of a Bi/Sb film with $T_{c0}=.272$ K in zero field at 60 mK and 130 mK. Similar IVs were observed in a sample farther from the SIT, $T_{c0}=.356$ K. At several distinct currents, large voltage jumps (singularities in *dV/dI*) occur. A 500 Ω series resistor was used to measure the current in the film, permitting the drops in current as a nonzero resistance appears in the film. The voltage jumps and the associated hysteresis (not shown) are reproducible. The structure appears below the temperature $T_0 \sim 140$ mK where the film's resistance is immeasurably small and presumably where Cooper pairs are globally condensed. Each IV curve displays a critical current asymmetry depending on the direction of the current ramp with the decreasing current arm noticeably smoother. At 60 mK, the larger value of the critical current density $J_c=60$ A/cm$^2$ (and $J_c=150$ A/cm$^2$ in a $T_{c0}=.356$ K Bi/Sb film at 70 mK). Above $T_0$, the IVs become linear at zero bias and non-hysteretic.

We associate the growth in the amplitude fluctuation dominated portion of the resistive transitions and the appearance of multiple critical currents with proximity of a disordered superconductor to a quantum critical point. While a simple Ginzburg criterion implies that the size of the critical region of a thermal phase transition grows with increasing disorder, it cannot account for the rate and character of the growth shown in Fig. 2. It predicts a disorder dependence of the transition width $\Delta T/T_{c0} \sim .5\, R_N/R_Q$ such that $\Delta T/T_{c0}$ linearly approaches .5 for a film at critical disorder $R_N = R_Q$ [8]. In previous work, the width of the transition $\Delta T/T_{c0}$ in films far from the SIT was consistent with the Ginzburg criterion [1,8,22]. This approach, however, neither extends to higher disorder nor accounts for the observed divergent transition width near the SIT. On the other hand, modeling the SIT as a continuous QPT, the vanishing energy scale at the quantum critical point is most naturally $T_{c0}$ or the condensation energy in a coherence volume, $U_{coh}$, in homogeneous films. The diverging transition width corresponds to an increase in the size of the regime for which $kT > U_{coh}$ as $U_{coh}$ becomes very small.

A recent theory [14] of the superconductor to normal metal quantum phase transition produces the functional dependence of the divergence of the transition width shown in Fig. 2. The model predicts quantum fluctuations in the critical temperature and consequently, growth in fluctuations of $\Delta$ upon approaching this quantum phase transition. The rms fluctuations in the transition temperature follow $\Delta T/T_{c0} \sim ln(T_1/T_{c0})$ in 2D homogeneously disordered films where $T_1$ is the characteristic temperature of the excitations that mediate the superconductivity (e.g. $T_1 \cong 300K$ for Pb). Though the theory is intended for 3D S-N transition, its prediction semi-quantitatively accounts for broadening in the temperature dependence of tunneling measurements on homogeneous PbBi/Ge films with $T_{c0}>1.7$ K [17]. While our transport data, which are from films much closer to the 2D SIT, are consistent with a logarithmic dependence, the divergence is faster than predicted, implying $T_1 \approx 4$ K. Differences in the sensitivity of transport and tunneling experiments to fluctuations in $T_{c0}$ or strong disorder effects, not accounted for by the theory, may be responsible for this discrepancy.

The anomalies in the zero field IVs suggest that amplitude fluctuations persist well below $T_{c0}$ in films close to the SIT, i.e. in the superconducting regime of the phase diagram, outside the regime in which the QCF dominate the physics. While reminiscent of the phase slip center IVs observed in Type I whiskers and films [23], the reentrance of this behavior in the strongly Type II limit on the verge of the QPT is not expected. The voltage jumps imply that the order parameter becomes disjunct at several independent points in the film at different critical superfluid velocities, that is, the film exhibits many spatially local critical currents. Thus, the order parameter amplitude or Cooper pair density must be non-uniform near the QCP. These fluctuations might correspond to the fluctuations in $T_{c0}$ that have been proposed to occur near the superconductor to normal metal QPT [14]. Alternatively, they may reflect the effects of random strong disorder in a homogeneous media which has been argued to create superconducting "blobs" within a normal state matrix [15,24,25]. Similar structure in IVs has been reported recently near the disorder tuned SIT in short quasi-1D granular Sn wires [26]. The inhomogeneities responsible in that case, however, are related to the "granular morphology" [27] of the wires rather than inhomogeneities that develop within a structurally homogeneous system.

We have assembled a consistent picture of the QCR of the SIT tuned by homogeneous disorder. The divergence of the amplitude fluctuation dominated portion of the transition suggests that disorder suppresses the order parameter amplitude to zero rather than localizing it. The latter picture would be consistent with the experimental evidence of the transition in granular systems. We picture the fluctuations in the quantum critical regime to occur between a ground state with Cooper pairs and one without. Once $kT < \Delta$, the superconducting state emerges as the Cooper pair wavefunctions become phase coherent. The inhomogeneity is then a function of the spatial or temporal quantum fluctuations of the amplitude which, in the very low superfluid density limit, form areas of normal material that coexist with the regions of condensate. Whether the $T=0$ QPT exclusively involves the growth of fluctuations in the phases of the superconducting order parameters of different regions in the film or also fluctuations in the order parameter amplitude remains an open issue [8,28]. The data presented here strongly suggest, however, that fluctuations in the amplitude of the order parameter are likely to dominate the experimentally accessible regime of the QPT.

We acknowledge helpful discussions with Brad Marston, Dietrich Belitz, Nandini Trivedi, and Xinsheng Ling and early contributions by Shih-ying Hsu. We are grateful for support from NSF grants DMR-9801983 and DMR-9502920.

not be accounted for by weak localization and electron-electron interaction effects. An effect this small does not influence our analysis or conclusions.

[21] Similarly, a logarithmic divergence is observed when $R_N(T)$ is taken to be transport measured at $H_{c2}$ at which the superconducting film's R(T) first exhibits agreement with the conductance of a WL insulator [$G \sim \log(T)$ at low T] to temperatures well below $T_{c0}$. $H_{c2}$, defined in this way, is found to be approximately two Tesla per Kelvin of $T_{c0}$ for films up to $T_{c0} \sim 3K$.

[22] Homogeneous films in the 1D limit exhibit excess broadening for $T_{c0}>1.6$ K (F. Sharifi, A. Herzog, and R. C. Dynes. Phys. Rev. Lett.**71**, 428 (1993)). This behavior is consistent with the expectation that quantum critical phenomena emerge farther from the critical point in lower dimensional systems.

[23] M. Skocpol, M. Beasley, and M. Tinkham, J. Low. Temp. Phys. **16**, 145 (1974).

[24] L. Bushalevskii, S. Panyukov, and M. Saovskii, Sov. Pyhs. JETP **65**, 380 (1987).

[25] D. Kowal and Z. Ovadyahu. Solid State Comm. **90**, 783 (1994).

[26] A. Frydman, E. P. Price, and R. C. Dynes, Solid State. Comm. **106**, 715 (1998).

[27] K. Ekinci and J. M. Valles, Jr. Phys. Rev. B. *To be published* (1998).

[28] N. Markovic, C. Christiansen, and A. M. Goldman, cond-mat/9808176.

**Figure Captions:**

Fig. 1 Electrical transport properties as a function of temperature of Bi/Sb films with thicknesses ranging from .7 to .9 nm near the disorder tuned SIT. a)Sheet resistance. b)Sheet conductance measured relative to the sheet conductance at 8K. Notice that the sheet conductances converge to a similar temperature dependence at high temperatures. c)Sheet resistance normalized by the temperature dependent normal state resistance.

Fig. 2 Normalized width of the superconducting transitions of Bi/Sb films as a function of $T_{c0}/T_c(bulk)$, where $\delta T/T_{c0} \equiv [T(.9R_N)-T(.1R_N)]/T(.5R_N)$ and $T_c(bulk)=6$ K is the transition temperature of a thick, low sheet resistance, Bi/Sb film. The dashed straight line is a guide to the eye.

Fig. 3  Current-voltage characteristics of a Bi/Sb film very near to the SIT, ($T_{c0}$=.275 K) at the two temperatures indicated.  Note the discontinuities in the 60 mK trace.

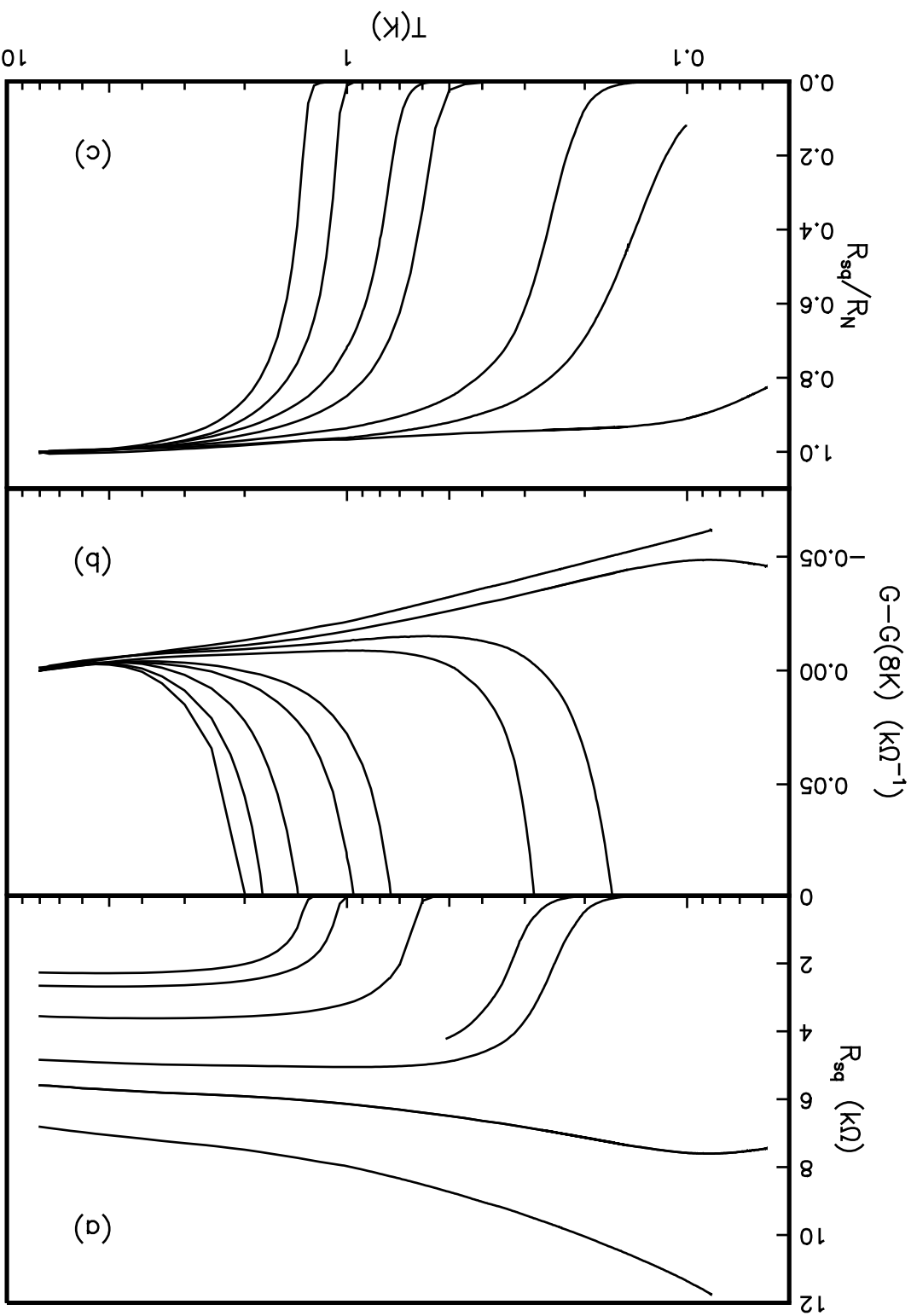

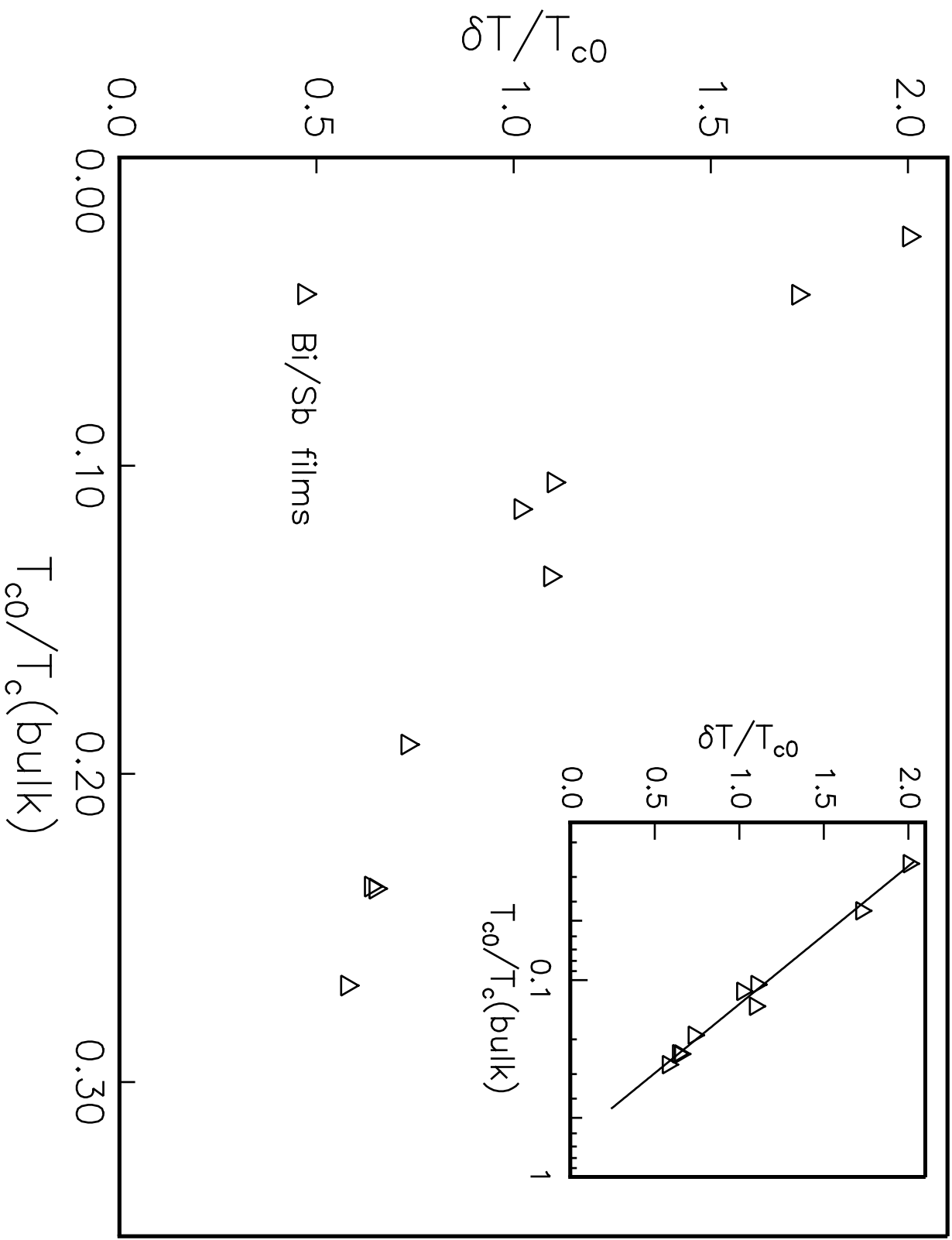

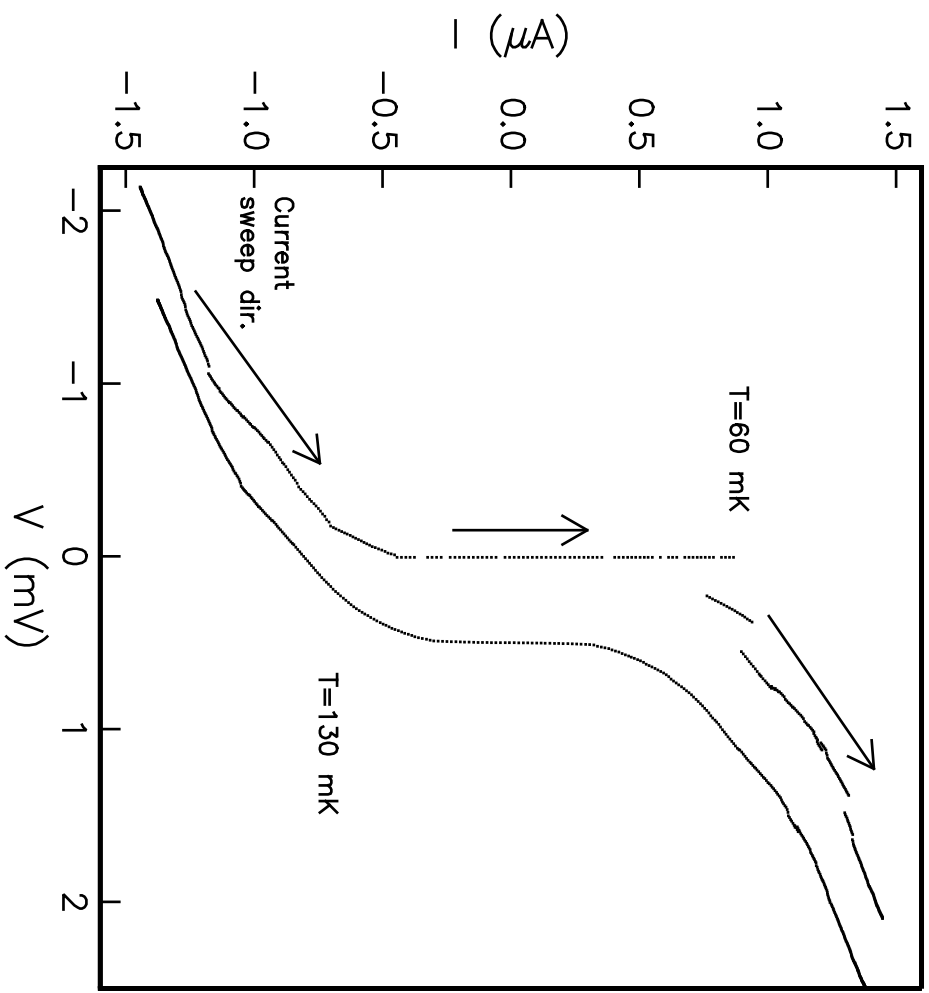